%% file: SUS-12-011_temp.tex
\pdfoutput=1

\documentclass[11pt,twoside,a4paper,cmspaper,final,collab]{cms-tdr}
\def\svnVersion{167924}\def\cmsCernNo{2012-184}\def\cmsCernDate{\today}\def\cmsMessage{Submitted to Physical Review Letters}

\begin{document}\cmsNoteHeader{SUS-12-011}

\hyphenation{had-ron-i-za-tion}
\hyphenation{cal-or-i-me-ter}
\hyphenation{de-vices}

\RCS$Revision: 149191 $
\RCS$HeadURL: svn+ssh://svn.cern.ch/reps/tdr2/papers/SUS-12-011/trunk/SUS-12-011.tex $
\RCS$Id: SUS-12-011.tex 149191 2012-09-26 03:29:07Z hatake $
\newcommand{\fulllumi}{4.98\xspace}
\newcommand{\squark}{\sQua}
\newcommand{\gluino}{\sGlu}
\newcommand{\mht}{\ensuremath{\slash\mkern-12mu{H}_{\text{T}}}\xspace}
\newcommand{\MHT}{\mht}
\newcommand{\ppbar}{\Pp\Pap\xspace}
\newcommand{\pp}{\Pp\Pp\xspace}
\newcommand{\e}{\Pe\xspace}
\newcommand{\epm}{\ensuremath{\Pe^\pm}\xspace}
\newcommand{\W}{\PW\xspace} 
\newcommand{\zmumu}{\ensuremath{\Z \rightarrow \mu^{+} \mu^{-}}\xspace}
\newcommand{\znunu}{\ensuremath{\Z \rightarrow \cPgn \cPagn}\xspace}
\newcommand{\zmumubr}  {\ensuremath{\Z (\mu^{+} \mu^{-})}\xspace}
\newcommand{\znunubr}{\ensuremath{\Z ( \cPgn\cPagn )}\xspace}
\newcommand{\zellellbr}{\ensuremath{\Z ( \ell^{+} \ell^{-} )}\xspace}
\newcommand{\wlnubr}{\ensuremath{\W (\ell \nu )}\xspace}
\newcommand{\wtaunubr} {\ensuremath{\W (\tau \nu )}\xspace}
\newcommand{\MEt}{\ETslash}
\providecommand{\tauh}{\ensuremath{\tau_\mathrm{h}}\xspace}
\renewcommand{\GeVc}{\GeV}
\renewcommand{\GeVcc}{\GeV}
\renewcommand{\TeVc}{\TeV}
\renewcommand{\TeVcc}{\TeV}
\newlength{\cmsFigWidth}
\ifthenelse{\boolean{cms@external}}{\setlength{\cmsFigWidth}{1.0\columnwidth}}{\setlength{\cmsFigWidth}{0.80\textwidth}}
\newlength{\cmsFigWidthII}
\ifthenelse{\boolean{cms@external}}{\setlength{\cmsFigWidthII}{1.0\columnwidth}}{\setlength{\cmsFigWidthII}{0.70\textwidth}}
\newlength{\cmsFigWidthApdx}
\ifthenelse{\boolean{cms@external}}{\setlength{\cmsFigWidthApdx}{0.90\textwidth}}{\setlength{\cmsFigWidthApdx}{0.90\textwidth}}
\cmsNoteHeader{SUS-12-011} 
\title{Search for new physics in the multijet and missing transverse momentum final state in proton-proton collisions at \texorpdfstring{$\sqrt{s}=7\TeV$}{sqrt(s) = 7 TeV}}

\author{The CMS Collaboration}

\date{\today}

\abstract{
A search for physics beyond the standard model
is performed in events with at least three jets and large missing transverse momentum
produced in proton-proton collisions at a center-of-mass energy of $\sqrt{s}=7\TeV$.
No significant excess of events above the expected backgrounds is observed in
4.98\fbinv of data collected with the CMS detector at the Large Hadron Collider.
The results are presented
in the context of the constrained minimal supersymmetric extension of the standard model
and more generically for simplified models. 
For the simplified models of gluino-gluino and squark-squark production, 
gluino masses below 1.0\TeVcc and squark masses below 0.76\TeVcc are excluded 
in case the lightest supersymmetric particle mass is below 200\GeV.
These results significantly extend previous searches.
}

\hypersetup{%
pdfauthor={CMS Collaboration},%
pdftitle={Search for new physics in the multijet and missing transverse momentum final state in proton-proton collisions at sqrt(s) = 7 TeV},%
pdfsubject={CMS},%
pdfkeywords={CMS, physics, supersymmetry}}

\maketitle 

\label{sec:introduction}

Many extensions of
the standard model (SM) of particle physics have been proposed
to address the shortcomings of the SM, e.g., problems concerning the gauge hierarchy
and identity of dark matter~\cite{SUSY0,LittleHiggs,UED}.
Supersymmetry (SUSY) is one such new physics model, 
which postulates
a new symmetry that relates fermionic and bosonic degrees of freedom
and introduces a superpartner for each SM particle.
In $R$-parity conserving models~\cite{FarrarEtAl}, SUSY particles are produced in pairs,
and the lightest SUSY particle (LSP) is stable.
If the LSP is weakly interacting and neutral, it
serves as a candidate for dark matter.
At the Large Hadron Collider (LHC),
squarks (\squark) and gluinos (\gluino), the superpartners of the quarks and gluons,
would be produced via the strong interaction and decay to SM particles and two LSPs.
A typical signature is the all-hadronic final state, characterized by
multiple jets arising from quarks and gluons,
and large missing transverse momentum
due to the unobserved LSPs.

Searches in this final state have been performed by
experiments at the Fermilab Tevatron~\cite{CDFLimits,D0Limits}
and at the LHC~\cite{RA2,RA1,RA1_2011,RAZOR,ATLASJetMET,ATLAS2011,ATLAS2011HighNjets,ATLAS2011FullHighNjets,ATLAS2011Full}.
This Letter presents 
a search in events with multiple jets and large missing transverse momentum
produced in 7\TeV \pp{} collisions using a data sample
corresponding to an integrated luminosity of $\fulllumi\pm 0.11 \fbinv$~\cite{lumi} collected
with the Compact Muon Solenoid (CMS) detector.
The search strategy follows Ref.~\cite{RA2} but uses more than 100 times
the amount of data.
This search is not specifically optimized for a particular SUSY model but is sensitive
to a variety of new physics models that lead to the multijet final state with large missing transverse momentum.
The results of this 
search are interpreted in the context of the constrained minimal supersymmetric extension of the SM
(CMSSM)~\cite{Chamseddine:1982jx,Arnowitt:1992aq,CMSSM}
and
in a more general context for simplified models~\cite{Alwall:2008ag,SMS-LHC-NP-Searches}
of new particles decaying to one or two jets and a stable weakly interacting particle.

\label{sec:detector}

The central feature of the CMS detector~\cite{CMS} is a superconducting solenoid of
6\unit{m} internal diameter, providing a magnetic field of 3.8\unit{T}.
Within the field volume are the silicon pixel and strip tracker,
the lead-tungstate crystal electromagnetic calorimeter,
and the brass and scintillator hadron calorimeter.
Charged particles 
are measured by the silicon tracker, covering $0 < \phi < 2\pi$ in azimuth and
$|\eta| < 2.5$~\cite{CoordiateSystem}.
The calorimeters 
surrounding the tracking volume cover 
$|\eta| < 3$.
Outside the field
the quartz and steel forward hadron calorimeters extend the coverage to $|\eta| < 5$.
Muons are identified in gas ionization detectors, covering $|\eta| < 2.4$, embedded in the steel return yoke of the magnet.
A two-tier trigger system selects the \pp collision events for use in this search.

The recorded events are reconstructed using the particle-flow algorithm~\cite{PFT-09-001},
which reconstructs particles, 
namely charged hadrons, photons, neutral hadrons, muons, and electrons,
using the information from all subdetectors.
These particles
are then clustered into jets
using the anti-$k_\mathrm{T}$ clustering algorithm
with distance parameter 0.5~\cite{antikt}.
Corrections are applied to account for the dependence of the jet response on transverse momentum
\pt and $\eta$~\cite{JETJINST}
and for the effects of additional (pileup) \pp{} collisions overlapping
with the collision of interest~\cite{PU_JET_AREAS,JET_AREAS}.

\label{sec:eventselection}

The event sample for the search is selected by requiring at least three jets
with 
$\pt>50\GeVc$ and $|\eta|<2.5$.
The further selection is based on two variables:
\HT{}, defined as
$\HT=\sum\pt$ where the sum is carried out over jets
with $\pt>50\GeVc$ and $|\eta|<2.5$, and
$\vec{\MHT}$, defined as $\vec{\MHT}=-\sum\vec\pt$
where the sum is over jets with $\pt>30\GeVc$ and $|\eta|<5$.
Events are required to have $\HT>500$\GeV and $\text{\MHT}>200$\GeV,
where \text{\MHT} is the magnitude of the $\vec{\MHT}$.
The \text{\MHT} requirement rejects
most of the QCD multijet background.
Events with $\vec{\MHT}$ aligned in azimuth
with one of the two leading jets with $\Delta\phi<0.5$ rad
or along the third jet with $\Delta\phi<0.3$ rad
are removed to further reduce the QCD multijet background.
Events containing isolated muons or electrons with $\pt>10\GeVc$ are also vetoed
in order to reject \ttbar and $\W/\Z$+jets backgrounds with leptons in the final state~\cite{MUO-10-002,EGM-10-004,RA2}.
Events are also rejected if a jet with $\pt>30\GeVc$ 
has an electromagnetic \pt fraction larger than 0.95 or a neutral hadron \pt fraction larger than 0.90.
In addition, events affected by instrumental effects,
particles from noncollision sources,
and poor reconstruction quality are rejected (event cleaning)~\cite{RA2,METJINST}.
All these requirements constitute the baseline selection~\cite{SupplementalMaterial}.
The event sample used in this search is collected by triggering on both \HT and \text{\MHT} 
or only on \HT{}. 
The \HT threshold ranges from 160 to 350 \GeV{}, and the \text{\MHT} threshold ranges from 60 to 110 \GeV{}.
The trigger efficiency is measured to be consistent with 100\% for the baseline event selection. 

To increase the sensitivity of the search to the different kinematic regions of signal events,
the sample of 1885 events passing the baseline selection
is divided into 14
subsamples defined in terms of the \HT and \text{\MHT} values (search selections),
as listed in the first column of Table~\ref{tab:FinalEventYields}.

The SM backgrounds 
mainly consist of
$\znunubr$+jets events and
$\wlnubr$+jets events from $\W$ or $\ttbar$ production ($\ell=\Pe$, $\mu$, or $\tau$).
The $\wlnubr$+jets events pass the search selection when
the $\Pe$/$\mu$ escapes detection 
or a $\tau$ decays hadronically. 
The QCD multijet events also contribute to the background when 
leptonic decays of heavy-flavor hadrons inside jets or jet energy mismeasurements lead to a large \text{\MHT}{}.
The contributions from other SM processes are found to be negligible.
In this search
all of the backgrounds are estimated from data~\cite{RA2}.

Several Monte Carlo (MC) samples are used
to model the signal as well as to develop and validate the background prediction methods.
The \ttbar, \W{}/\Z{}+jets, and $\gamma$+jets samples are produced using the 
\MADGRAPH{}5~\cite{madgraph} generator, interfaced with the \PYTHIA{} 6.4.24~\cite{pythia}
parton-shower model.
The \ttbar and \W{}/\Z{}+jets samples are 
scaled up to the next-to-leading-order (NLO) or next-to-next-to-leading-order
cross section predictions~\cite{Kidonakis:2010dk,Melnikov:2006kv}.
The QCD multijet 
and SUSY signal production is simulated with \PYTHIA{} 6.4.24,
the CTEQ6L~\cite{CTEQ6} parton distribution functions (PDFs),
and a CMS custom underlying event tuning~\cite{pythia-z2}.
The generated events are passed through a
\GEANTfour-based~\cite{Geant4-2} detector simulation, and
have the same distribution of pileup \pp{} interactions
as observed in the data.

\label{sec:bg}

The $\znunubr$+jets background contribution is
estimated using $\gamma$+jets events by treating photons as \znunu decays.
The \Z boson and photon exhibit similar kinematic properties at high \pt{}, and the hadronic component of events is
similar in the two cases \cite{Bern:2011pa,Kuhn:2005gv,Ask:2011xf,Bern:2012vx}.
A $\gamma$+jets sample is collected by triggering on a $\gamma$ candidate
with or without an additional requirement on \HT{} depending on the data-taking period.
The photon candidates~\cite{CMS-PAS-QCD-10-037} are required to be isolated from other particles in
the tracker and calorimeters and to have the shower shape consistent with that for a prompt photon.
In order to predict the $\znunubr$+jets background, 
the $\gamma$+jets sample is corrected for
the $\gamma$ reconstruction efficiency
and purity, both measured from data \cite{RA2}, and
the $\znunubr$+jets/$\gamma$+jets production ratio, obtained from the \MADGRAPH{} simulation samples,
which also takes into account the detector acceptance for photons.
The total multiplicative correction factor to obtain the $\znunubr$+jets background prediction
from the $\gamma$+jets event yield
is 0.28$\pm$0.06 for the baseline selection.
The dominant systematic uncertainties on this background estimation 
originate from the theoretical uncertainty on
the $\gamma$/\Z{} cross section ratio~(20--40\%)~\cite{Bern:2011pa,Bern:2012vx},
the detector acceptance~(5--7\%),
and the $\gamma$ reconstruction and isolation efficiency~(1--10\%), depending on the search regions.

As a cross check, the \znunubr{}+jets background is also estimated using \zmumubr{}+jets events by
treating muons as neutrinos
and correcting for the acceptance and efficiencies of the \zmumubr{}+jets event selection
and the ratio of branching fractions $\mathcal{B}(\znunu)/\mathcal{B}(\zmumu)=5.95\pm0.02$~\cite{PDG}.
The \znunubr{}+jets background estimated with this method is found to be consistent with the one from the
$\gamma$+jets events.

\begin{table*}[t!]
\fontsize{9 pt}{1.1 em}
\selectfont
\begin{center}
\topcaption{Event yields for different backgrounds
for the 14 search selections together with the total backgrounds,
as determined from the collision data, and number of events observed in data.
The quoted uncertainties are the combinations of the statistical and systematic uncertainties.
}
\label{tab:FinalEventYields}
{
\begin{tabular}{|rr|rl|rl|rl|rl|rl|r|}
\hline

\multicolumn{2}{|c|}{Selection}
            & \multicolumn{2}{c|}{$\znunu$}
            & \multicolumn{2}{c|}{$\ttbar/\W$}
            & \multicolumn{2}{c|}{$\ttbar/\W$}
	    & \multicolumn{2}{c|}{QCD}
            & \multicolumn{2}{c|}{Total}
	    & \multicolumn{1}{c|}{~~Data~~}          \\

\HT (GeV) &\MHT (GeV)    & \multicolumn{2}{c|}{}
	    & \multicolumn{2}{c|}{$\to \e,\mu+\cmsSymbolFace{X}$}
            & \multicolumn{2}{c|}{$\to \tauh{}+\cmsSymbolFace{X}$}
	    & \multicolumn{2}{c|}{multijet}
            & \multicolumn{2}{c|}{background}
	    & \multicolumn{1}{c|}{}  \\
\hline

500--800    &200--350  & 359   &$\pm$ 81      & 327    &$\pm$ 47      &  349   &$\pm$ 40      & 119    &$\pm$ 77          & 1154   &$\pm$ 128     & 1269 \\
500--800    &350--500  & 112   &$\pm$ 26      &  48    &$\pm$  9      &   62.5 &$\pm$  8.7    &   2.2  &$\pm$  2.2        &  225   &$\pm$  29     &  236 \\
500--800    &500--600  &  17.6 &$\pm$  4.9    &   5.0  &$\pm$  2.2    &    8.7 &$\pm$  2.5    &   0.0  &$\pm$  0.1        &   31.3 &$\pm$   5.9   &   22 \\
500--800    &$>$600    &   5.5 &$\pm$  2.6    &   0.8  &$\pm$  0.8    &    2.0 &$\pm$  1.8    &   0.0  &$\pm$  0.0        &    8.3 &$\pm$   3.2   &    6 \\
\hline
800--1000   &200--350  &  48   &$\pm$ 19      &  58    &$\pm$ 15      &   56.3 &$\pm$  8.3    &  35    &$\pm$ 24          &  197   &$\pm$  35     &  177 \\
800--1000   &350--500  &  16.0 &$\pm$  6.7    &   5.4  &$\pm$  2.3    &    7.2 &$\pm$  2.0    &   1.2  &$^{+1.3}_{-1.2}$  &   29.8 &$\pm$   7.5   &   24 \\
800--1000   &500--600  &   7.1 &$\pm$  3.7    &   2.4  &$\pm$  1.5    &    1.3 &$\pm$  0.6    &   0.0  &$^{+0.2}_{\phantom{+}0.0}$  &   10.8 &$\pm$   4.0   &    6 \\
800--1000   &$>$600    &   3.3 &$\pm$  1.7    &   0.7  &$\pm$  0.7    &    1.0 &$\pm$  0.3    &   0.0  &$^{+0.1}_{\phantom{+}0.0}$  &    5.0 &$\pm$   1.9   &    5 \\
\hline
1000--1200  &200--350  &  10.9 &$\pm$  5.1    &  13.7  &$\pm$  3.8    &   21.9 &$\pm$  4.6    &  19.7  &$\pm$ 13.3        &   66   &$\pm$  15     &   71 \\
1000--1200  &350--500  &   5.5 &$\pm$  3.0    &   5.0  &$\pm$  4.4    &    2.9 &$\pm$  1.3    &   0.4  &$^{+0.7}_{-0.4}$  &   13.8 &$\pm$   5.5   &   12 \\
1000--1200  &$>$500    &   2.2 &$\pm$  1.7    &   1.6  &$\pm$  1.2    &    2.3 &$\pm$  1.0    &   0.0  &$^{+0.2}_{\phantom{+}0.0}$   &    6.1 &$\pm$   2.3   &    4 \\
\hline
1200--1400  &200--350  &   3.1 &$\pm$  1.8    &   4.2  &$\pm$  2.1    &    6.2 &$\pm$  1.8    &  11.7  &$\pm$  8.3        &   25.2 &$\pm$   8.9   &   29 \\
1200--1400  &$>$350    &   2.3 &$\pm$  1.5    &   2.3  &$\pm$  1.4    &    0.6 &$^{+0.8}_{-0.6}$  &   0.2  &$^{+0.6}_{-0.2}$ & 5.4 &$\pm$   2.3   &    8 \\
\hline
$>$1400     &$>$200    &   3.2 &$\pm$  1.8    &   2.7  &$\pm$  1.6    &    1.1 &$\pm$  0.5    &  12.0  &$\pm$  9.1        &   19.0 &$\pm$   9.4   &   16 \\
\hline

\end{tabular}
}
\end{center}
\end{table*}

The $\wlnubr$+jets events ($\ell=\Pe$ or $\mu$) from \W or top quark production constitute
a background when an electron or muon is not identified or is nonisolated and therefore
passes the lepton veto.
This 
background is estimated from a $\mu$+jets control sample,
selected with the same criteria
as those used for the search except that we require
exactly one rather than zero isolated $\mu$.
The transverse mass
$m_\mathrm{T}=\sqrt{2\pt^{\mu}\MEt[1-\cos(\Delta\phi)]}$ is required to be
less than 100\GeVcc
in order to select events containing a $\W\to\mu\nu$ decay and to suppress
possible new physics signal contamination, i.e.,
the signal events resulting in the $\mu$+jets sample used for the background estimation.
Here \text{$\MEt{}$} is the missing transverse energy~\cite{METJINST},
and $\Delta\phi$ is the azimuthal angle between the $\mu$ and the \text{$\MEt{}$}.
Events are weighted according to
$\left(1/\epsilon_\text{iso}^{\mu}\right)
[(1-\epsilon_\text{reco}^{\e,\mu})/\epsilon_\text{reco}^{\mu}]$
and
$\left(\epsilon_\text{reco}^{\e,\mu}/\epsilon_\text{reco}^{\mu}\right)
[(1-\epsilon_\text{iso}^{\e,\mu})/\epsilon_\text{iso}^{\mu}]$
in order to predict events with unidentified leptons and nonisolated leptons, where
$\epsilon_\text{reco}^{\e,\mu}$ and $\epsilon_\text{iso}^{\e,\mu}$
are the reconstruction and isolation efficiencies of the electrons and muons.
The lepton reconstruction efficiencies are obtained from MC simulation, while
the isolation efficiencies are extracted by applying a ``tag-and-probe'' method~\cite{EWK-10-005}
on the \zellellbr{}+jets events in data.
The lepton reconstruction and identification efficiencies are parametrized
in lepton \pt and $\Delta R=\sqrt{(\Delta\eta)^2+(\Delta\phi)^2}$
relative to the closest jet,
in order to account for the kinematic differences between \zellellbr{}+jets events
and the \ttbar and \W{}+jets events.
Leptons that are out of acceptance
and events lost due to the $m_\mathrm{T}$ requirement
are accounted for using factors determined from simulation.
This background estimation method 
based on the collision data
is validated by applying it to a MC sample and
comparing the predicted and the true detector-level background distributions.

The predicted 
background for each search region is listed in Table~\ref{tab:FinalEventYields}.
On this background estimation,
low statistics in the $\mu$+jets
control sample are
the dominant source of uncertainty
in most of the search regions.
The modeling of the lepton reconstruction and isolation efficiencies
yields a 10\% uncertainty.
An additional uncertainty of 4--20\% varying for different search regions
is assigned based on the statistical power of the validation
of this background estimation method.
A 3\% uncertainty accounts for the effect of the presence of QCD, \Z{},
or diboson events in the $\mu$+jets sample, which are modeled by MC simulation.

The background from the hadronic decay of $\tau$ leptons (\tauh)
is estimated from a sample of $\mu$+jets events,
selected from inclusive $\mu$ or $\mu+{\geq}2$-jet triggers by requiring
exactly one $\mu$ with $\pt>20\GeVc$ and $|\eta|<2.1$.
In this sample, the muon \pt is replaced with a jet \pt taken randomly
from a simulated response function
for a hadronically-decaying $\tau$ lepton.
The \HT{} and \text{\MHT} of the event are recalculated including this $\tau$ jet,
and the search selections are applied to predict the \tauh background.
The $\tau$-jet response function for $\pt^{\text{jet}}/\pt^\tau$
is obtained from simulated $\ttbar$ and $\wtaunubr$+jets events
by matching the reconstructed $\tau$ jet
with the generated $\tau$.
Corrections are applied to account for the trigger efficiency,
acceptance, and efficiency of the $\mu$ selection, and
the ratio of branching fractions
$\mathcal{B}(\W\rightarrow\tauh\nu)/\mathcal{B}(\W\rightarrow\mu\nu) = 0.69 \pm 0.05$~\cite{PDG}.
This \tauh background estimation method is validated by applying it to the \W and \ttbar MC samples,
and 6--13\% uncertainties are assigned mainly to reflect the statistical power of this validation.
The other main systematic uncertainties 
arise from the $\mu$ acceptance ($\leq$13\%);
the $\tau$-jet response function ($\leq$20\%);
and the subtraction of residual QCD multijet, \zmumubr{}+jets,
and $(\ttbar / \W) \to \tau\nu{}+\cmsSymbolFace{X}\to \mu\nu{}+\cmsSymbolFace{X}$ backgrounds ($\leq$2\%),
where the quoted uncertainties apply to all search regions.

The QCD background is estimated from collision data~\cite{RA2} recorded with a set of
triggers having an \HT threshold ranging from 150 to 700\GeV.
The data samples used include the electroweak contributions not removed by the lepton
veto and any potential new physics events; however, their cross section is
negligible compared to the QCD multijet cross section.
First, the \pt values of the jets with $\pt>15\GeV$ in these events are adjusted
within the jet \pt resolution, using a kinematic fit such that the events are
balanced in the transverse plane.
The jet \pt values in
the rebalanced events are then smeared with the measured jet resolutions
to predict the QCD multijet background.
The jet \pt response functions are determined as a function of
\pt and $\eta$ using a 
QCD multijet MC sample
that includes heavy-flavor quarks.
The width and tail of the \pt response functions are subsequently
adjusted to account for the differences in the resolutions measured
in simulation and in data~\cite{JETJINST}.
The width ($\sigma$) of the Gaussian part of the simulated response is 5 (30)\% narrower than
what is observed in the data for $|\eta|<0.5$ $(2.3<|\eta|<5.0)$.
After correcting for this difference,
the fraction of jets with response more than $2.5\sigma$ away from the mean value
is consistent with that in the data within uncertainties.
The main uncertainties in this QCD estimation method arise from the shape of the jet response functions
including the Gaussian width, the tails, the heavy-flavor contribution,
and the effect of pileup on jets in an event.
The method has been validated in simulated QCD multijet events 
within the statistical uncertainties 
(30--50\%), 
which are assigned as an additional uncertainty.
The total uncertainty adds up to 60--70\%.

\label{sec:sensitivity}

\begin{figure}[th!]
\begin{center}
\includegraphics[width=\cmsFigWidth]{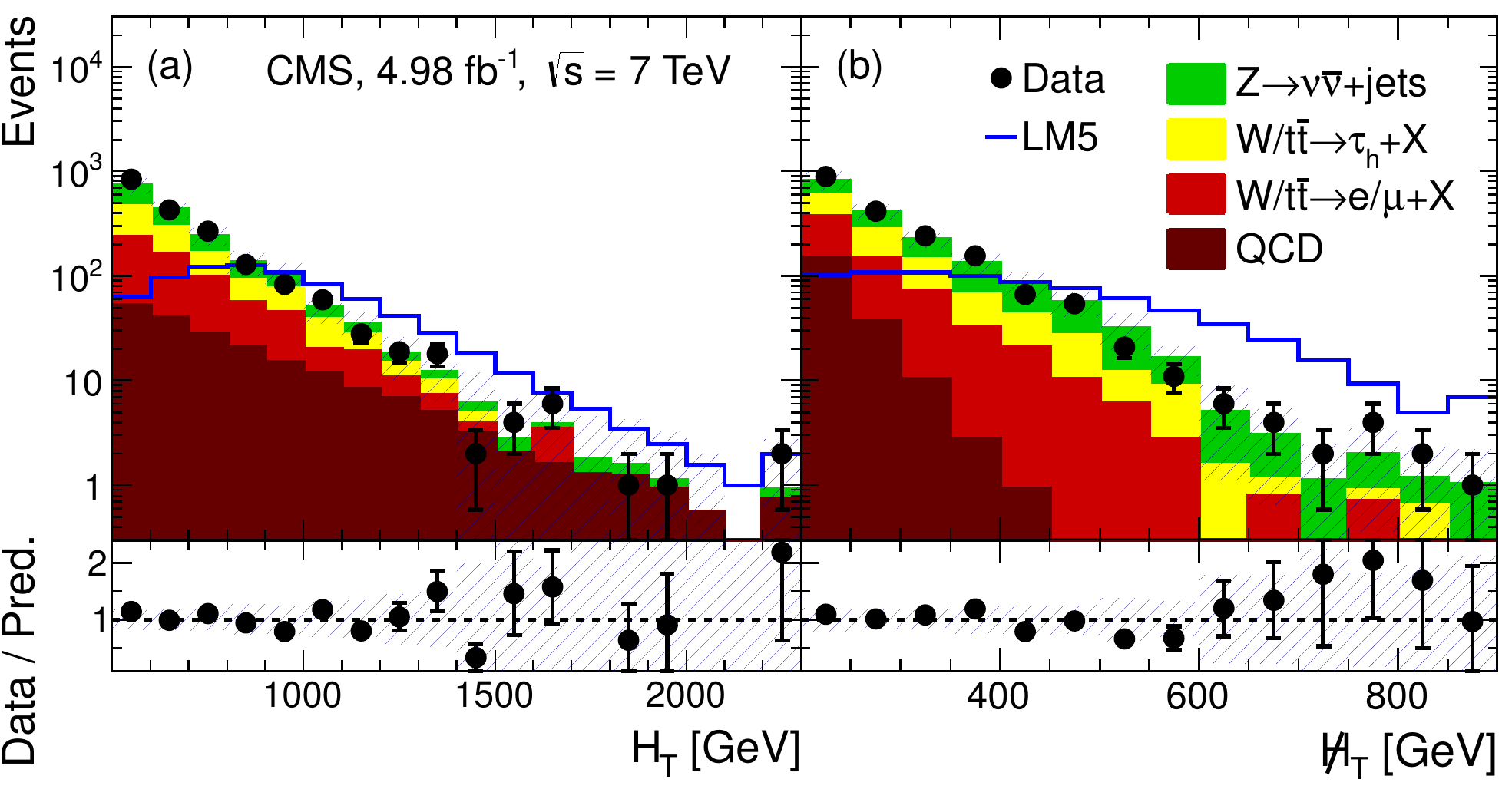}
\end{center}
\caption{The (a) \HT and (b) \MHT distributions in the search data samples (circles)
compared with histograms showing predictions of the SM background and SUSY signal
(LM5, see the text) for events passing the baseline selection.
The hatched region indicates the uncertainties on the background predictions.
The last bin contains all events above the maximum values of \HT{} and \MHT{} in the figures.
The ratio of the observed data to the background predictions is also shown.
}
\label{fig:htmhtcombined}
\end{figure}

The predicted yields of the SM background and the number of events observed in data
are summarized in Table~\ref{tab:FinalEventYields} for the 14 search regions. 
Figure~\ref{fig:htmhtcombined} shows the \HT and \MHT distributions
predicted for the
SM background, together with those observed in data.
The data are consistent with the SM background estimates.

\begin{figure}[thb!]
\begin{center}
\includegraphics[width=\cmsFigWidthII]{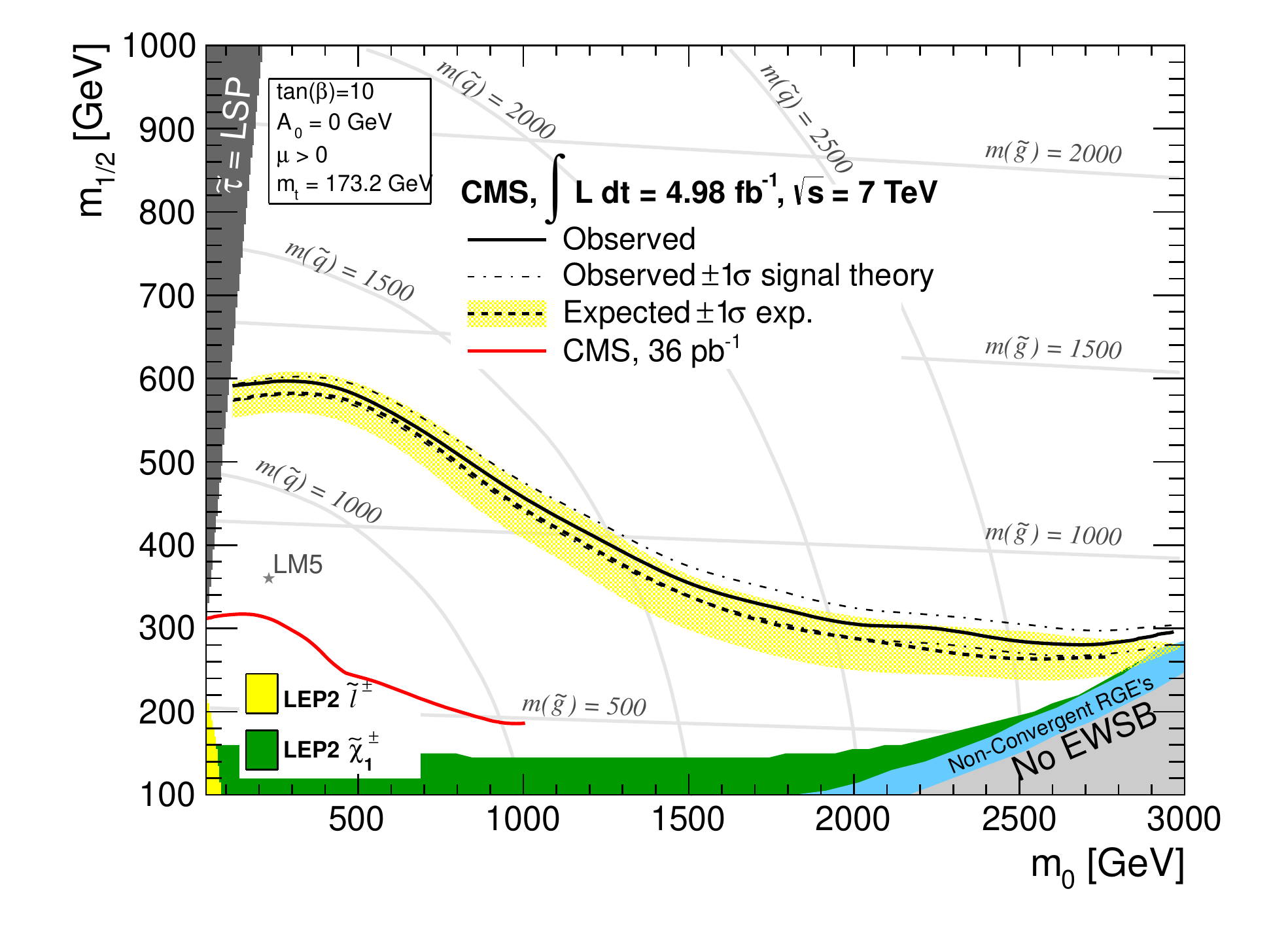}
\end{center}
\caption{The observed and expected $95\%$ C.L. limits in the CMSSM ($m_0$, $m_{1/2}$) plane.
The yellow-shaded region shows the $\pm1\sigma$ variation in the expected limit,
while  the dot-dashed curves show the variation in the observed limit
when the signal cross section is varied by its theoretical uncertainties.
The remaining CMSSM parameters are $\tan\beta=10$, $\mu>0$, and $A_0=0$.
The limits from an earlier CMS search~\cite{RA2} and from other experiments~\cite{LEPLimits} are also shown.
The limits from Ref.~\cite{RA2} are shown only up to 1000 \GeVcc in $m_0$, as done in \cite{RA2}.
The regions where the superpartner of the $\tau$ lepton ($\tilde\tau$) is the LSP, the
renormalization group equations (RGE's) do not converge, or there is no electroweak
symmetry breaking (EWSB)~[53], are also indicated.
}
\label{fig:msugralimits}
\end{figure}

The 95\% confidence level (C.L.) upper limits on the CMSSM signal cross section 
are set using a modified frequentist CL$_\mathrm{s}$ method,
taking the profile likelihood as a test statistic~\cite{aread,tjunk,ATL-PHYS-PUB-2011-011}.
The results from 14 exclusive search regions are combined into one test statistic considering the
bin-to-bin correlations of the systematic uncertainties.
The CMSSM model has
five independent parameters:
the 
universal scalar and gaugino masses
at the grand unification scale,
$m_0$ and $m_{1/2}$;
the trilinear coupling, $A_0$;
the ratio of the vacuum expectation values of the two Higgs doublets, $\tan\beta$;
and the sign of the Higgsino mixing parameter, $\mu$.
The signal cross section is calculated at NLO and next-to-leading-log (NLL) accuracy~\cite{SUSYNLO,SUSY-resummation,LHC-SUSY-Xsec}.
The \HT and \MHT distributions predicted for a low-mass CMSSM benchmark parameter set LM5,
$m_0= 230\GeVcc$, $m_{1/2}= 360\GeVcc$, $A_0= 0$, $\tan\beta = 10$, and $\mu>0$,
are shown in Fig.~\ref{fig:htmhtcombined}.

The acceptance times efficiency of the event selection for signal events 
is evaluated using the simulated CMSSM samples.
The uncertainties on the background predictions,
the luminosity determination (2.2\%)~\cite{lumi},
the signal acceptance and efficiency arising from the jet energy correction (8\%),
the jet energy resolution (2\%), 
the PDF (6\%),
the trigger inefficiency (2\%),
and the event cleaning~\cite{METJINST} (3\%)
are taken into account by the limit-setting procedure.
The possible overprediction of the backgrounds due to the presence of the signal in the data samples
used for the background
prediction is estimated to be about 3--20\%, depending on ($m_0$, $m_{1/2}$) values,
and subtracted when testing for the signal+background hypothesis
in the CL$_\mathrm{s}$ method.

The upper limits on the CMSSM signal cross section are mapped into lower limits
in the ($m_0$, $m_{1/2}$) plane (exclusion contour),
as shown in Fig.~\ref{fig:msugralimits}~\cite{CMSSM_template,SupplementalMaterial}.
The exclusion contours are also shown
for the cases in which the signal cross section is varied
by changing the renormalization and factorization scales by a factor of 2 
and using the PDF4LHC recommendation~\cite{PDF4LHC} for the PDF uncertainty
to illustrate the sensitivity of the exclusion to the signal cross section
uncertainty.
Conservatively, using the $-1\sigma$ theory uncertainty values on the observed limit,
squark masses below 1.2\TeVcc and
gluino masses below 0.72\TeVcc are excluded
for the chosen CMSSM parameter set.

\begin{figure}[t!]
\begin{center}
\includegraphics[width=\cmsFigWidth]{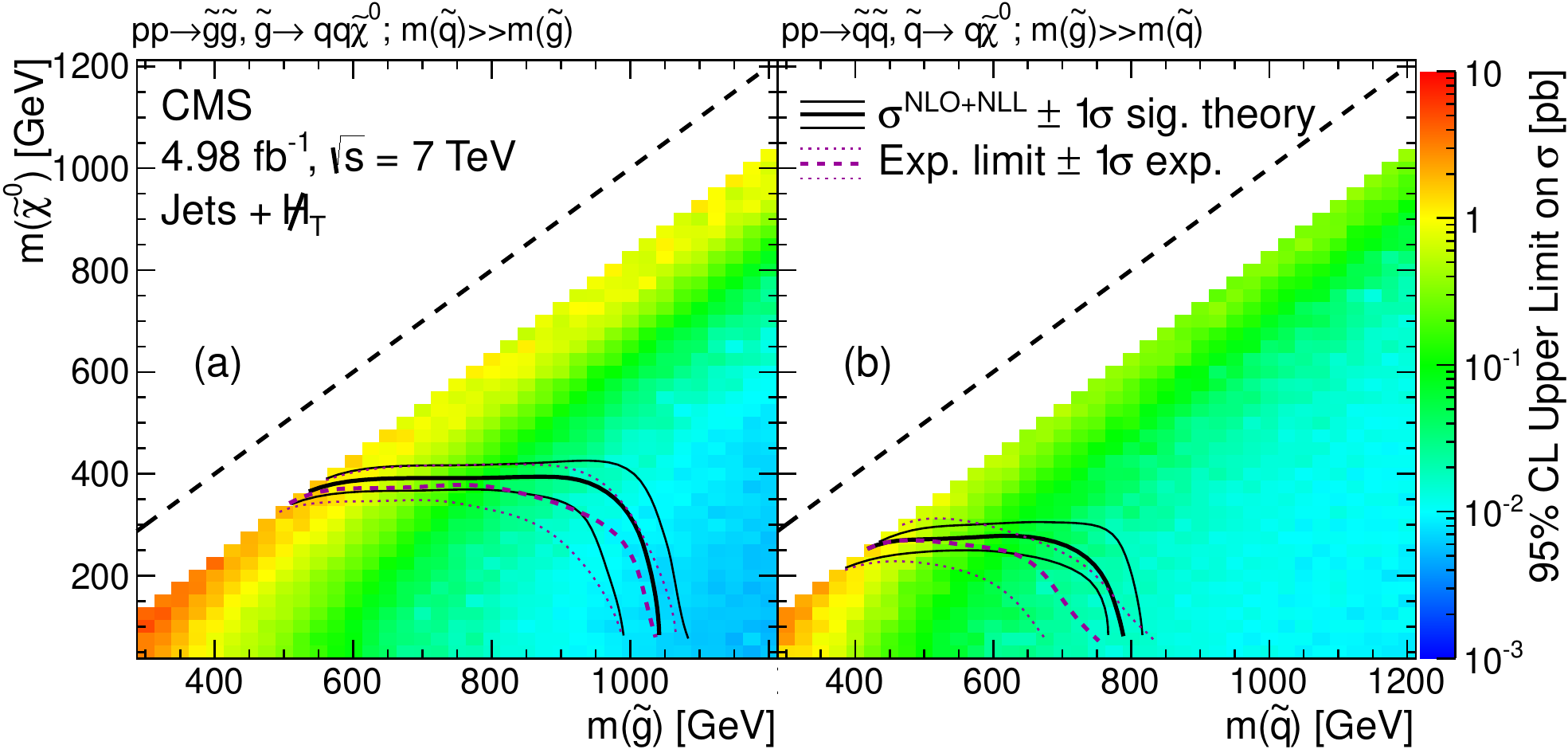}
\end{center}
\caption{The observed and expected 95\% C.L. upper limits on the
(a) $\gluino\gluino$ and
(b) $\squark\squark$ cross sections
in the $\left(m(\gluino), m(\chiz)\right)$ and $\left(m(\squark), m(\chiz)\right)$ planes
obtained with the simplified model.
Also shown are the $\pm1\sigma$ variation in the expected limit and
the variation in the observed limit when the signal cross section is varied by its
theoretical uncertainties.
}
\label{fig:smslimits}
\end{figure}

The search results are also presented in a more general context of simplified models~\cite{Alwall:2008ag,SMS-LHC-NP-Searches}
of new particles (\squark or \gluino) decaying to one or two jets and an undetectable weakly interacting particle ($\chiz$).
The model used here includes the production of $\gluino\gluino$ and $\squark\squark$ pairs and their decays
for a wide range of $\left(m(\gluino), m(\chiz)\right)$ and $\left(m(\squark), m(\chiz)\right)$ values,
and other SUSY particles are decoupled by being given masses beyond the reach of the LHC.
The signal acceptance times efficiency~\cite{SupplementalMaterial} and its uncertainty
are evaluated in the same way as used for the CMSSM but using the simulated simplified model signal samples.
The observed and expected $95\%$ C.L. upper limits on the signal cross section of
$\gluino\gluino$ and $\squark\squark$
production are shown in Fig.~\ref{fig:smslimits}
in the $\left(m(\gluino), m(\chiz)\right)$ and $\left(m(\squark), m(\chiz)\right)$ planes,
together with contours where the signal cross sections from the NLO+NLL
calculations~\cite{SUSYNLO,SUSY-resummation,LHC-SUSY-Xsec} are excluded.
The results are presented only in the region of $m(\gluino{},\squark)-m(\chiz)>150\GeVc$,
since the estimation of signal acceptance times efficiency becomes unreliable
due to its strong dependence on the modeling of QCD radiation
when the mass difference $m(\gluino{},\squark) - m(\chiz)$ is smaller.
In this model, the $m(\gluino)$ values below 1.0\TeVcc
and $m(\squark)$ values below 0.76\TeVcc are excluded for $m(\chiz)<200\GeVcc$.

\label{sec:conclusions}

In summary, a search for new physics has been performed
in the final state with at least three jets and large \MHT
using a data sample
corresponding to an integrated luminosity of $\fulllumi \fbinv$ collected in 7\TeV{}
\pp{} collisions 
with the CMS detector at the LHC.
The observed numbers of events are consistent with the estimated SM background contributions,
and 95\% C.L. exclusion limits are set in the CMSSM parameter space which
significantly extend the previous results. 
For the simplified models of $\gluino\gluino$ and $\squark\squark$ production,
the $m(\gluino)$ values below 1.0\TeVcc
and $m(\squark)$ values below 0.76\TeVcc are excluded for $m(\chiz)<200\GeVcc$.

\label{sec:ack}

We wish to congratulate our colleagues in the CERN accelerator
departments for the excellent performance of the LHC machine. We thank
the technical and administrative staff at CERN and other CMS
institutes and acknowledge support from FMSR (Austria); FNRS and FWO
(Belgium); CNPq, CAPES, FAPERJ, and FAPESP (Brazil); MES (Bulgaria);
CERN; CAS, MoST, and NSFC (China); COLCIENCIAS (Colombia); MSES
(Croatia); RPF (Cyprus); Academy of Sciences and NICPB (Estonia);
Academy of Finland, ME, and HIP (Finland); CEA and CNRS/IN2P3
(France); BMBF, DFG, and HGF (Germany); GSRT (Greece); OTKA and NKTH
(Hungary); DAE and DST (India); IPM (Iran); SFI (Ireland); INFN
(Italy); NRF and WCU (Korea); LAS (Lithuania); CINVESTAV, CONACYT,
SEP, and UASLP-FAI (Mexico); PAEC (Pakistan); SCSR (Poland); FCT
(Portugal); JINR (Armenia, Belarus, Georgia, Ukraine, Uzbekistan); MST
and MAE (Russia); MSTD (Serbia); MICINN and CPAN (Spain); Swiss
Funding Agencies (Switzerland); NSC (Taipei); TUBITAK and TAEK
(Turkey); STFC (United Kingdom); and DOE and NSF (USA).

\bibliography{auto_generated}   

\ifthenelse{\boolean{cms@external}}{}{
\clearpage
\appendix
\input{supplemental}}

\cleardoublepage \appendix\section{The CMS Collaboration \label{app:collab}}\begin{sloppypar}\hyphenpenalty=5000\widowpenalty=500\clubpenalty=5000\input{SUS-12-011-authorlist.tex}\end{sloppypar}
\end{document}

%% file: supplemental.tex
\section{Supplemental Material: Baseline and search event selections\label{sec:supplemental}}
\ifthenelse{\boolean{cms@external}}{\onecolumngrid}{}

The event selection starts from a baseline selection.
Events passing the baseline selection are then divided into 14 exclusive search regions.
The baseline selection requirements after trigger are:

\begin{itemize}
\item at least three jets with $\pt> 50\GeVc$ and $| \eta | < 2.5$;
\item $\HT > 500\GeV$; 
\item $\mht> 200\GeV$;  
\item $ | \Delta\phi ( J_{n} , \vec{\mht} ) | > 0.5$ rad, $n=1,2$ and $ | \Delta\phi ( J_{3} , \vec{\mht} ) | > 0.3$ rad,
  where $\Delta\phi$ is the azimuthal angle difference between jet axis $J_n$ and the $\vec{\mht}$ direction for the three
  highest-\pt{} jets in the event;
\item no isolated muons or electrons in the event;
  \begin{itemize}
  \item muons and electrons are required to have $\pt\ge 10$\GeVc and a good quality track that is matched
  to the primary vertex within $200\mum$ transversely and $1\cm$ longitudinally;
  \item they are required to be isolated, with a relative isolation variable, defined as
    ${\left[ \sum^{\Delta R<0.3}\pt{}^{\text{charged hadron}} +\sum^{\Delta R<0.3}\pt{}^{\text{neutral hadron}} +\sum^{\Delta R<0.3}\pt{}^\text{photons} \right] } / {\pt}$,
    smaller than 0.2, where
    $\pt{}^{\text{charged hadron}}$,
    $\pt{}^{\text{neutral hadron}}$, and
    $\pt{}^\text{photons}$ are, respectively, the transverse momenta of charged hadrons, neutral hadrons,
    and photons, as reconstructed by the particle-flow algorithm, within a distance $\Delta R=0.3$ in $\eta$-$\phi$ space
    of the lepton;
  \item muons are required to have $|\eta|<2.4$,
    whereas electrons should have $|\eta| < 2.5$ excluding the barrel-endcap transition region $1.44 < |\eta| < 1.57$;
  \end{itemize}
\item jets with $\pt>30\GeVc$
  have an electromagnetic \pt fraction less than 0.95
  and a neutral hadron \pt fraction less than 0.90.
\end{itemize}
Events passing the baseline selection are divided into 14 search regions:
\begin{itemize}
\item for the \HT bins of 500--800 and 800--1000\GeV, \MHT is binned
into 200--350, 350--500, 500--600, and $>$600\GeV;
\item for the \HT bin of 1000--1200\GeV, \MHT is binned into 200--350, 350--500, and $>$500\GeV;
\item for the \HT bin of 1200--1400\GeV, \MHT is binned into 200--350\GeV and $>$350\GeV;
\item for the \HT bin of \HT{}$>$1400\GeV, \MHT{}$>$200\GeV.
\end{itemize}

\begin{figure*}[p!]
\begin{center}
\includegraphics[width=0.90\textwidth]{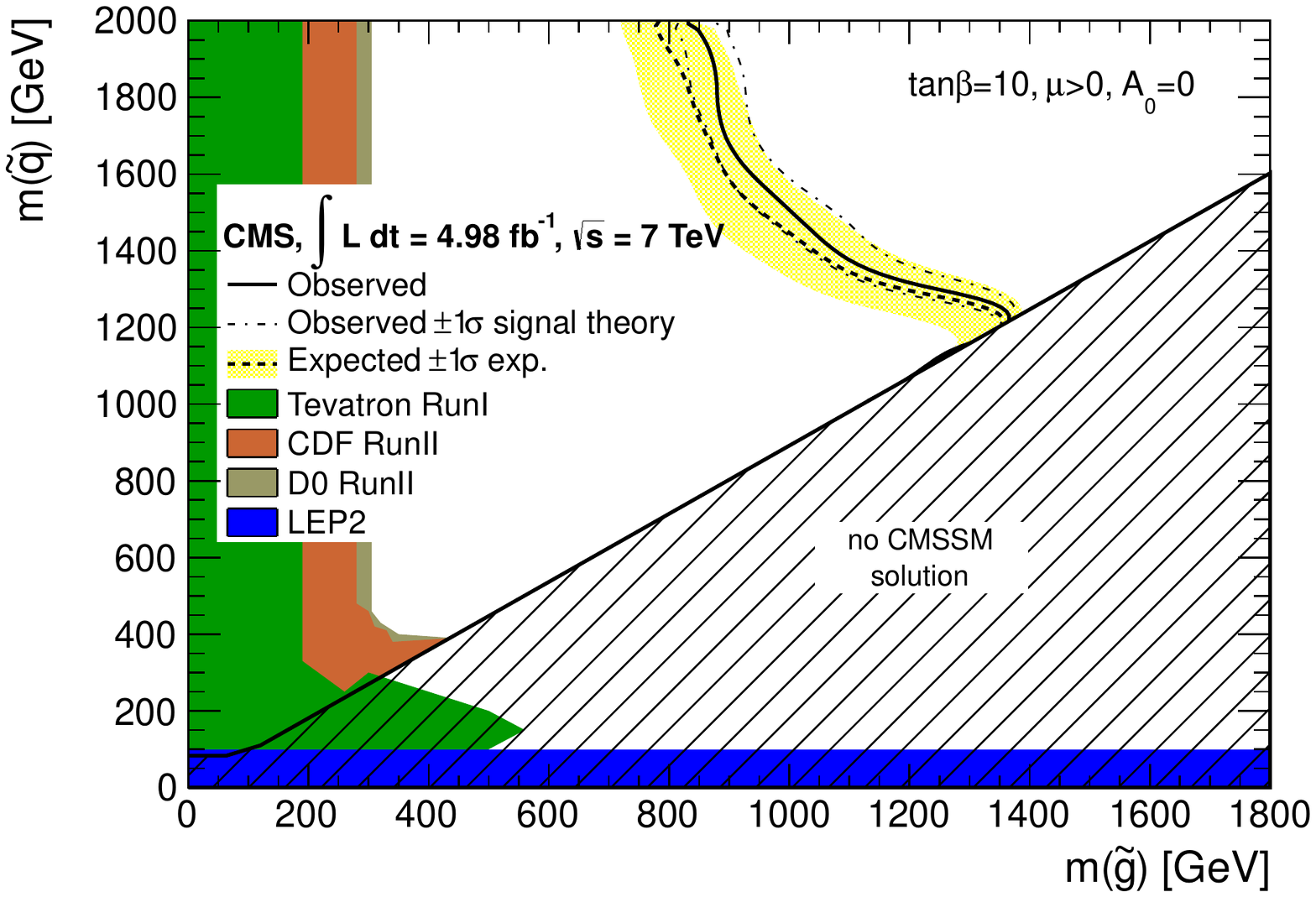}
\end{center}
\caption{The observed and expected 95\% CL lower limits in the CMSSM
$\left(m(\gluino), m(\squark)\right)$ plane, for
$\tan\beta=10$, $\mu>0$, and $A_0=0$.
The yellow-shaded region shows the $\pm1\sigma$ variation in the expected limit, while the dot-dashed
curves show the variation in the observed limit when the signal cross section is varied by its
theoretical uncertainties. The limits from earlier searches by other experiments~\cite{CDFLimits,D0Limits,LEPLimits} are also shown.
Comparisons with earlier searches are shown for illustrative purpose only,
as they are derived with different models or parameter choices.
}
\label{fig:msugralimits2}
\end{figure*}

\begin{figure*}[p!]
\begin{center}
\includegraphics[width=\cmsFigWidthApdx]{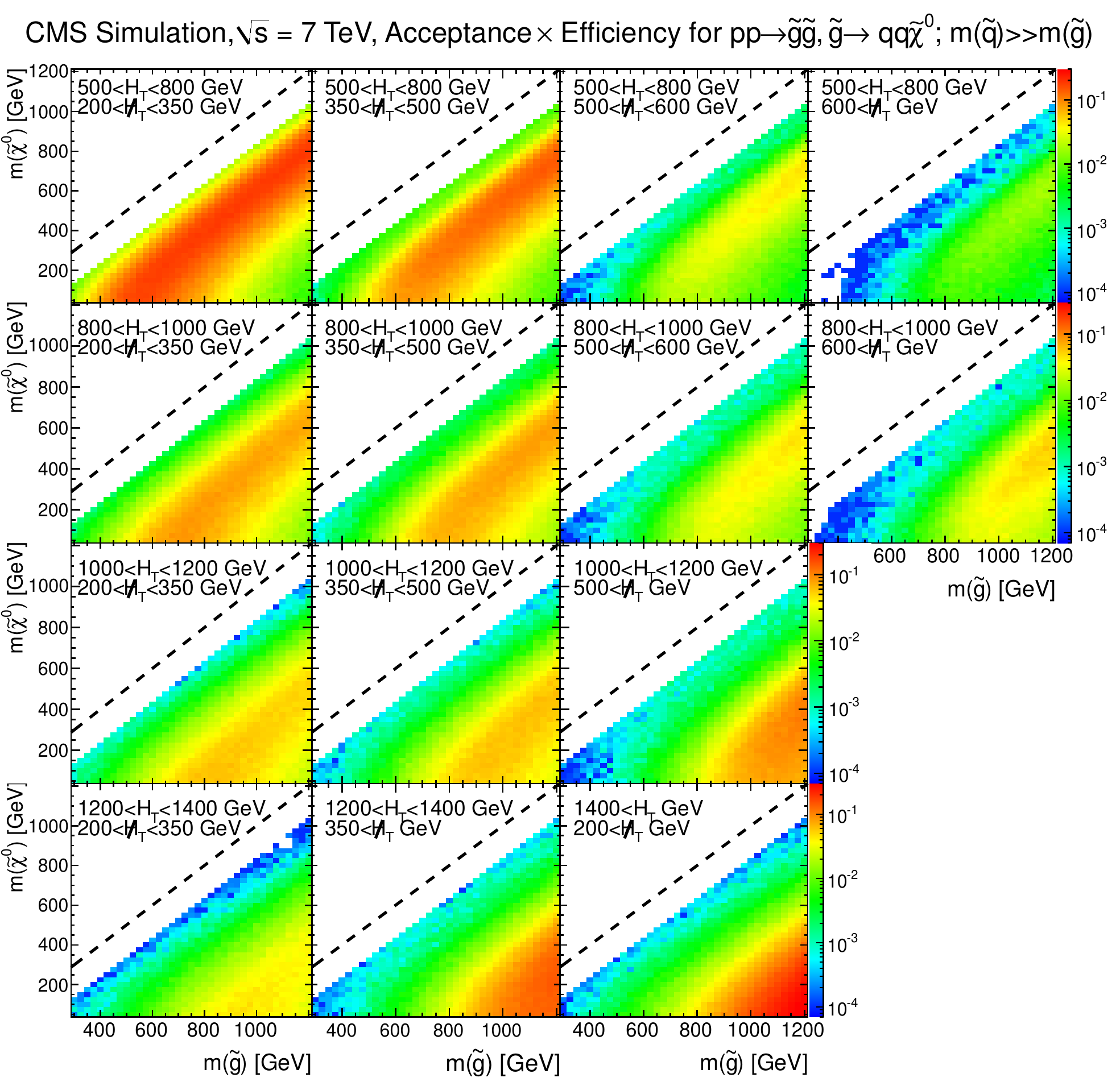}
\end{center}
\caption{The acceptance times efficiency of the 14 search regions
for the simplified model of $\gluino\gluino$ ($\gluino\to \cmsSymbolFace{q}\cmsSymbolFace{q}\chiz$) production
in the $\left(m(\gluino), m(\chiz)\right)$ plane.
Empty points are due to the low acceptance times efficiency where
no simulated signal events pass the search selection.
}
\label{fig:sms_efficiency_t1}
\end{figure*}

\begin{figure}[p!]
\begin{center}
\includegraphics[width=\cmsFigWidthApdx]{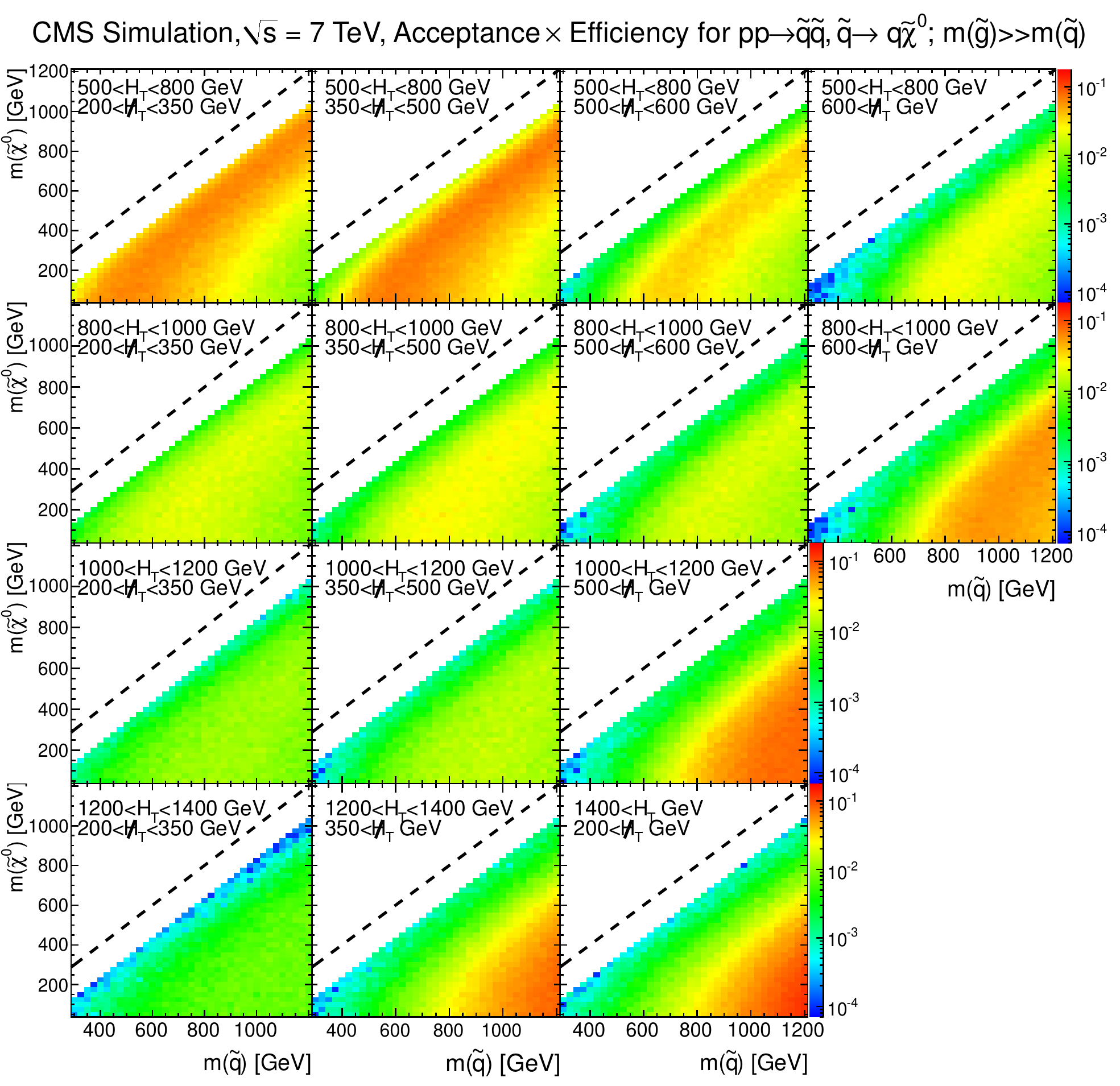}
\end{center}
\caption{The acceptance times efficiency of the 14 search regions
for the simplified model of $\squark\squark$ ($\squark\to \cmsSymbolFace{q}\chiz$) production
in the $\left(m(\squark), m(\chiz)\right)$ plane.
Empty points are due to the low acceptance times efficiency where
no simulated signal events pass the search selection.
}
\label{fig:sms_efficiency_t2}
\end{figure}

%% file: SUS-12-011-authorlist.tex
\textbf{Yerevan Physics Institute,  Yerevan,  Armenia}\\*[0pt]
S.~Chatrchyan, V.~Khachatryan, A.M.~Sirunyan, A.~Tumasyan
\vskip\cmsinstskip
\textbf{Institut f\"{u}r Hochenergiephysik der OeAW,  Wien,  Austria}\\*[0pt]
W.~Adam, T.~Bergauer, M.~Dragicevic, J.~Er\"{o}, C.~Fabjan\cmsAuthorMark{1}, M.~Friedl, R.~Fr\"{u}hwirth\cmsAuthorMark{1}, V.M.~Ghete, J.~Hammer, N.~H\"{o}rmann, J.~Hrubec, M.~Jeitler\cmsAuthorMark{1}, W.~Kiesenhofer, V.~Kn\"{u}nz, M.~Krammer\cmsAuthorMark{1}, D.~Liko, I.~Mikulec, M.~Pernicka$^{\textrm{\dag}}$, B.~Rahbaran, C.~Rohringer, H.~Rohringer, R.~Sch\"{o}fbeck, J.~Strauss, A.~Taurok, P.~Wagner, W.~Waltenberger, G.~Walzel, E.~Widl, C.-E.~Wulz\cmsAuthorMark{1}
% ...
\vskip\cmsinstskip
\dag:~Deceased\\
1:~~Also at Vienna University of Technology, Vienna, Austria\\
% ...


%% file: SUS-12-011_temp.bbl
\begin{thebibliography}{10}
\bibitem{CoordiateSystem}
\hrefCMSnoop {} {}CMS uses a right-handed coordinate system, with the origin at
  the nominal interaction point, the $x$ axis pointing to the center of the
  LHC, the $y$ axis pointing up (perpendicular to the LHC plane), and the $z$
  axis along the anticlockwise-beam direction. The polar angle, $\theta$, is
  measured from the positive $z$ axis, and the azimuthal angle, $\phi$, is
  measured in the $x$-$y$ plane relative to the $x$ axis. The pseudorapidity
  $\eta$ is defined as $\eta = -\ln[ \tan (\theta/2)]$.
\end{thebibliography}
